\def \SAIT #1 #2 {{\em Mem.\ Soc.\ Astron.\ It.\/} {\bf #1}, #2}
\def \MESS #1 #2 {{\em The Messenger\/} {\bf #1}, #2}
\def \ASTRNACH #1 #2 {{\em Astron. Nach.\/} {\bf #1}, #2}
\def \AAP #1 #2 {{\em Astron. Astrophys.\/} {\bf #1}, #2}
\def \AAL #1 #2 {{\em Astron. Astrophys. Lett.\/} {\bf #1}, L#2}
\def \AAR #1 #2 {{\em Astron. Astrophys. Rev.\/} {\bf #1}, #2}
\def \AAS #1 #2 {{\em Astron. Astrophys. Suppl. Ser.\/} {\bf #1}, #2}
\def \AJ #1 #2 {{\em Astron. J.\/} {\bf #1}, #2}
\def \ANNREV #1 #2 {{\em Ann. Rev. Astron. Astrophys.\/} {\bf #1}, #2}
\def \APJ #1 #2 {{\em Astrophys. J.\/} {\bf #1}, #2}
\def \APJL #1 #2 {{\em Astrophys.. J. Lett.\/} {\bf #1}, L#2}
\def \APJS #1 #2 {{\em Astrophys. J. Suppl.\/} {\bf #1}, #2}
\def \APSS #1 #2 {{\em Astrophys. Space Sci.\/} {\bf #1}, #2}
\def \ASR #1 #2 {{\em Adv. Space Res.\/} {\bf #1}, #2}
\def \BAIC #1 #2 {{\em Bull. Astron. Inst. Czechosl.\/} {\bf #1}, #2}
\def \JSQRT #1 #2 {{\em J. Quant. Spectrosc. Radiat. Transfer\/} {\bf #1}, #2}
\def \MN #1 #2 {{\em Mon. Not. R. Astr. Soc.\/} {\bf #1}, #2}
\def \MEM #1 #2 {{\em Mem. R. Astr. Soc.\/} {\bf #1}, #2}
\def \PLR #1 #2 {{\em Phys. Lett. Rev.\/} {\bf #1}, #2}
\def \PASJ #1 #2 {{\em Publ. Astron. Soc. Japan\/} {\bf #1}, #2}
\def \PASP #1 #2 {{\em Publ. Astr. Soc. Pacific\/} {\bf #1}, #2}
\def \NAT #1 #2 {{\em Nature\/} {\bf #1}, #2}

\documentstyle{memsait}
\input epsf.sty
\begin{opening}
\title{FORMATION OF SPHEROIDAL GALAXIES: THE SUB-MM VIEW}
\author{G. De Zotti$^1$, G.L. Granato$^1$, C. Baccigalupi$^2$,
M. Magliocchetti$^2$, P. Mazzei$^1$, F. Perrotta$^{1,2}$, L.
Danese$^2$}
\institute{$^1$Osservatorio Astronomico, Padova, Italy\\
$^2$SISSA, Trieste, Italy}
\date{} 
\end{opening}

    \def\lsim{\, \lower2truept\hbox{${< \atop\hbox{\raise4truept\hbox{$\sim$}}}$}\,}
\def\gsim{\, \lower2truept\hbox{${> \atop\hbox{\raise4truept\hbox{$\sim$}}}$}\,}

\begin{document}

\oddpagefooter{}{}{} 
\evenpagefooter{}{}{} 
\
\bigskip

\begin{abstract}
The intensity of the Cosmic Far-IR Background and the strong
evolution of galaxies in the far-IR to mm wavelength range
demonstrate that the bulk of starlight emitted during the early
phases of galaxy evolution was reprocessed by dust. Therefore, the
optical view of the galaxy formation process is highly incomplete
and biased, and must be complemented with far-IR/sub-mm
observations. We review the impact of sub-mm surveys on our
understanding of the evolutionary history of spheroidal galaxies.
A recent model, bringing into play also the inter-relationships
between formation and early evolution of spheroidal galaxies and
quasars, is described and some implications are outlined.
\end{abstract}

\section{Introduction}
The last ten years have seen a crescendo of discoveries that have
revolutionized our understanding of the process by which the tiny
graininess seen in the Cosmic Microwave Background (CMB) resulted
in the formation of galaxies.

Up to a decade ago, galaxy evolution models followed the approach
pioneered by  Tinsley (1972), Larson \& Tinsley (1974), Tinsley \&
Gunn (1976), the evolutionary population synthesis approach,
wherein galaxies form at some redshift $z_f$, perhaps varying by
type, with some star formation history $\psi(t)$ and some stellar
initial mass function (IMF). The general framework was that
postulated by Eggen et al. (1962) and Sandage (1986): elliptical
galaxies had essentially a simple passive evolution after
formation at high $z$ in a single rapid collapse and starburst,
while the timescale for star formation of late type galaxies is
much longer and in fact comparable with the Hubble time. Models
were required to reproduce the photometric properties of local
galaxies. The more refined models dealt simultaneously with both
photometric and chemical evolution. Guiderdoni \& Rocca-Volmerange
(1987) were the first to introduce {\it dust absorption}. However,
the general attitude was that dust extinction is a mere correction
that does not change the main evolutionary trends seen in the
optical band. This view appeared to be grounded by the fact that a
comparison of the optical and far-IR luminosity densities in the
local universe shows that only $\simeq 30\%$ of starlight is
reprocessed by dust.

Mazzei et al. (1992, 1994) produced the first self-consistent
chemical and photometric evolution models including both dust
absorption and emission, to get complete spectral energy
distributions (SEDs) from UV to far-IR both for late- and
early-type galaxies. These models showed that, while the SEDs of
late-type galaxies are not expected to change drastically
throughout their evolutionary history, under plausible assumptions
spheroidal galaxies might have been very optically thick during
their early phases, if indeed most of their stars were formed in a
giant starburst. Their SEDs would then have been similar to that
of hyper-luminous IRAS galaxies. Mazzei \& De Zotti (1996) showed
that (sub)-mm observations of a number of high redshift radio
galaxies supported this scenario. These models were exploited by
Franceschini et al. (1994) to produce the first set of galaxy
counts at optical {\it and} far-IR to sub-mm wavelengths and to
predict the SED of the Cosmic IR Background (CIRB); a remarkable
result was that the CIRB intensity might be as large as, or larger
than, that of the optical background.

Up to that time, the optical quest for ``normal'' high-$z$
(``primeval'') galaxies yielded a very meager return: only a
handful of objects were discovered, mostly serendipitously or
through absorption features in the spectra of distant quasars or
gravitational lensing. The only effective method for finding
high-$z$ galaxies was optical identifications of radio sources,
which are generally associated to early type galaxies.

A major breakthrough occurred roughly at the same time, thanks to
the color selection technique targeting the Lyman discontinuity at
$912\,$\AA, developed by Steidel \& Hamilton (1992, 1993).
Follow-up spectroscopy with the Keck telescope confirmed the
efficiency of this technique (Steidel et al. 1996) which led to
the discovery of a widespread population of high-$z$ galaxies.
%

Madau et al. (1996) applied the same technique to the multicolor
images of the Hubble Deep Field to identify galaxies at $2.0 < z <
3.4$ ($U$ drop-outs) and at $3.5 < z < 4.5$ ($B$ drop-outs).  They
compared the luminosity density in HDF high-$z$ galaxy samples to
that found in low-$z$ survey and interpreted these data in terms
of evolution with redshift of the star-formation/metal-production
rate. Although the results at high-$z$ were presented as lower
limits (because nearly all the corrections to be applied drive the
derived star formation rates up) it was argued that they may be
not far from the true values and that they are consistent with a
peak in the metal production rate at $1< z <2$. There was a
remarkable, at least qualitative, agreement between the metal
production rate, $\dot{\rho}_Z(z)$, estimated in this way and the
prediction of the cosmic chemical evolution models by Pei \& Fall
(1995), and the predictions of hierarchical CDM models (Baugh et
al. 1998), all of which show a peak in the metal production rate
at $z \sim $1--2. On the other hand, their conclusions were
affected by several large uncertainties due to (Ferguson et al.
2000; Somerville et al. 2001):

\begin{enumerate}

\item Selection criteria: the adopted color selection was extremely
conservative and spectroscopic surveys (Steidel et al. 1996;
Lowental et al. 1997) have identified at least a dozen $2<z<3.5$
galaxies that would have been missed

\item Sampling of the luminosity function: Madau et al. (1996) simply
summed up the observed fluxes of galaxies, although only the
bright tail of the LF was sampled (Thompson et al. 2001).

\item Surface brightness corrections: for fixed luminosity and physical
size, the surface brightness drops by about 1 mag between $z=3$
and $z=4$. This may produce a substantial underestimate of
$\dot{\rho}_Z(z)$ at high $z$ (Ferguson 1998; Lanzetta et al.
1999).

\item Effects of dust: dust is ubiquitous in star-forming regions and
relatively modest amounts of dust suffice to strongly attenuate
the UV emission. Analysis of UV spectral slopes yielded evidences
of substantial dust extinction (Pettini et al. 1998; Meurer et al.
1997, 1999; Sawicki \& Yee 1998; Steidel et al. 1999; Flores et
al. 1999; Mobasher \& Mazzei 2000). The derived upward corrections
of the luminosity densities by Madau et al. (1996) range from
factors of $\sim 3$ to more than 10, indicating that the
correction based on optical/UV data is both large and very
uncertain. Even more: very dusty galaxies, akin to ultra-luminous
IRAS galaxies, that were present at high $z$, would be largely
unaccounted by these studies because they would be unlikely to be
identified as Lyman-break galaxies.

\end{enumerate}

\noindent All that boils down to the conclusion that the optical
view of ``primeval'' galaxies is incomplete and biased. It must be
complemented with far-IR/sub-mm observations.

Almost simultaneous with the discovery of LBGs was a second major
breakthrough: the discovery of the Cosmic Infrared Background
(Puget et al. 1996; Guiderdoni et al. 1997; Schlegel et al. 1998;
Fixsen et al. 1998; Hauser et al. 1998; Lagache et al. 1999, 2000;
Finkbeiner et al. 2000) at a level 10 times higher than the
no-evolution predictions based on the IRAS local luminosity
function of galaxies and twice as high as the Cosmic Optical
Background obtained from the optical counts, implying that most of
starlight emitted at early phases of galaxy evolution is actually
reprocessed by dust (remember that in the local universe dust
emission accounts for only $\simeq 30\%$ of the bolometric
luminosity of galaxies). The CIRB SED turned out to be in
remarkably good agreement with the models by Franceschini et al.
(1994) whereby most of the star formation in spheroidal galaxies
was enshrouded by dust. Thus, {\it far-IR to sub-mm observations
carry the most direct, minimally biased information on the cosmic
star-formation/metal-production history.}

Sub-mm observations have a particularly privileged role in the
game: their uniquely strong, negative K-correction, due to the
steep increase with frequency of dust emission in galaxies,
coupled with the strong cosmological evolution demonstrated by ISO
and SCUBA data (Elbaz et al. 1999; Smail et al. 1997; Hughes et
al. 1998; Barger et al. 1999a,b; Eales et al. 1999) as well as by
the CIRB intensity, greatly emphasizes high-redshift galaxies.
Such observations therefore hold the key to understanding the
formation and evolution of massive elliptical galaxies, a still
controversial issue in cosmology.

\section{Monolithic versus merging scenarios for the formation of spheroidal galaxies}
Two quite different scenarios are still confronting each other.
One scenario, motivated by hierarchical clustering Cold Dark
Matter (CDM) cosmologies, predicts that most large galaxies arise
from a series of merging events taking place over a major fraction
of the cosmological time (Baugh et al. 1998; Kauffmann 1996).

The other scenario assumes instead that the bulk of the baryonic
mass of the galaxy was assembled in gaseous form. Most stars
formed in a single gigantic starburst, followed by essentially
passive evolution, due to the ageing of stellar populations. This
scheme is sometimes qualified as {\it monolithic}.

It is worth stressing that the {\it monolithic} scenario too can
be consistent with the hierarchical gravitational instability
picture for the structure formation. The difference with the {\it
merging} scenario is that large ellipticals formed most of their
stars {\it in situ} as soon as the corresponding dark matter halos
condensed out. In the {\it merging} scenario most stars are formed
in smaller galaxies that subsequently merged and the merging
events eventually triggered further bursts of star formation.

In the {\it merging} hypothesis, ellipticals would exhibit a broad
range of colors and spectral properties, reflecting the variety of
star formation histories, the bulk of stars having an intermediate
age, and their number density should decrease with increasing
redshift. On the contrary, in the {\it monolithic} scenario the
number density should remain constant and the stars should be old
and essentially coeval.

Several lines of observational evidence (the tightness of the
fundamental plane for ellipticals in local clusters, the tight
color-magnitude relation for ellipticals in clusters up to $z \sim
1$, the modest shift with increasing redshift in the zero-point of
the fundamental plane, Mg--$\sigma$, and color-magnitude relations
for cluster ellipticals, the small zero-point offset between
cluster and field ellipticals) converge in indicating that most
stars in galactic spheroids are old (formed at $z\gsim 2$--3) and
essentially coeval (see Renzini \& Cimatti 1999, Shade et al.
1999, and Kodama et al. 1999 for recent discussions).

It is, however, still controversial whether ellipticals were
assembled to their present mass at a later epoch, compared to that
of formation of their stars (see Renzini \& Cimatti 1999, Ferguson
et al. 2000, and references therein, and the recent work by
Martini 2001 and Rodighiero et al. 2001). Good agreement was
recently found by Daddi et al. (2000) between the observed surface
density of Extremely Red Objects (EROs) in the widest field survey
for such objects available, with that predicted by Pure Luminosity
Evolution (PLE) models after applying the appropriate color and
luminosity thresholds. Since there are indications that the bulk
of EROs are passively evolving ellipticals, this result supports
the view that most field ellipticals were fully assembled by $z
\sim 2.5$.

\section{The role of sub-mm surveys}
Sub-mm data provide information on an earlier stage of formation
of ellipticals than optical data do, i.e. on the stage when these
galaxies were forming most of their stars. SCUBA surveys did in
fact provide the most critical test for the {\it merging}
scenario\footnote{We will not deal here with the parameterized
empirical approaches, evolving backwards in time the local
luminosity functions at different wavelengths. Successful models
of this kind were worked out by Franceschini et al. (1991),
Franceschini (2000),
Pearson \& Rowan-Robinson (1996), Dwek et al. (1998), Blain et al.
(1999), Tan et al. (1999), Rowan-Robinson (2001), Pearson et al.
(2001), Takeuchi et al. (2001).}.

A very comprehensive attempt to make detailed predictions for the
observable properties of galaxies throughout the electromagnetic
spectrum, at all redshifts, starting from an assumed initial
spectrum of density fluctuations, was carried out by Granato et
al. (2000). These authors combined the semi-analytical galaxy
formation model of Cole et al. (2000; GALFORM) with the
evolutionary spectro-photometric models including  the effect of
dust by Silva et al. (1998; GRASIL), to compute epoch-dependent
galaxy luminosity functions from UV to sub-mm wavelengths.

GALFORM evolves the primordial density fluctuations and applies
simplified analytical descriptions of the main physical processes
of gas cooling and collapse, star formation, feedback effect from
supernovae, galaxy merging, chemical enrichment of stars and gas,
and more, on top of a Monte Carlo description of the process of
formation and merging of dark matter halos through hierarchical
clustering.

GRASIL follows the evolution of stellar populations and absorption
and emission by dust with a realistic 3D geometry, with a disk and
a bulge, two-phase dust (in clouds and in the diffuse ISM), star
formation in the clouds, radiative transfer of starlight through
the dust distribution, a realistic dust grain model including PAHs
and quantum heating of small grains, and a direct evaluation of
the dust temperature distribution at each point in the galaxy.

The models include both galaxies forming stars quiescently in
disks, and starbursts triggered by galaxy mergers. They reproduce
quite well a large number of observables, including the SEDs of
normal spirals and starbusts from UV to sub-mm; their internal
extinction properties and, in particular, the observed
relationship between far-IR/UV luminosity ratio and the slope of
the UV continuum and the observed starbust attenuation law
(Calzetti et al. 2000) with the same dust mixture which reproduces
the Milky Way extinction law; the local galaxy luminosity
functions from UV to far-IR; the counts of galaxies in wavebands
from UV to far-IR.

{\it However, they fail by a factor $\sim 10$ to reproduce the
sub-mm counts}, which turn out to be the most critical test for
semi-analytic approach.

In fact, analogous conclusions are reached by  Devriendt \&
Guiderdoni (2000) who implemented a independent {\it ab-initio}
models, building on earlier work by Guiderdoni et al. (1997, 1998)
and exploiting the spectro-photometric evolution models, including
dust, developed by  Devriendt et al. (1999; STARDUST). To
reproduce the $850\,\mu$m counts they are forced to introduce an
{\it ad-hoc} population of heavily extinguished, massive galaxies
with large star-formation rates at intermediate and high
redshifts.

These difficulties, affecting even the best current recipes, may
indicate that new ingredients need to be taken into account. A key
new ingredient may be the mutual feedback between formation and
evolution of spheroidal galaxies and of active nuclei residing at
their centers.

\section{Relationships between quasar and galaxy formation}
In the past, evolutionary histories of quasars and of galaxies
have been dealt with as essentially independent subjects. This was
due to two main reasons. On one side, it was essentially
impossible to have a synoptic view of the evolutionary history of
the two populations, since the high redshift universe was almost
exclusively the realm of quasars while the ``local'' universe
($z<1$) was the realm of galaxies. On the other side, and perhaps
more important, physical processes involved in the formation and
evolution of both quasar and galaxies are of great complexity and
therefore difficult to handle without the guidance of direct
observational data.

The situation has drastically improved in the last years with the
detection of ordinary and starburst galaxies up to $z>3$ which has
opened to direct investigation the early phases of galaxy
evolution. There is know convincing evidence of substantial
star-formation activity at least up to $z\simeq 4$ (Lagache et al.
1999; Steidel et al. 1999).

On the other hand, extensive data on the demography of
super-massive black holes (BH) in nearby galaxies has been
accumulating, allowing to piece together an increasingly
quantitative description of their mass function (Salucci et al.
1999) and of where (and possibly when) they formed. There is no
evidence of quasar activity outside galaxies and deep imaging of
bright quasar samples have revealed that their hosts are most
likely elliptical galaxies (Hall \& Green 1998; McLure et al.
1999; Schade et al. 2000).

As recognized by Haehnelt \& Rees (1993), the epoch of quasar
activity coincides with the time when the first potential wells
form in the standard CDM scenario. In the same framework, Haehnelt
et al. (1998) demonstrated that the luminosity function of
actively star-forming galaxies at $z=3$ and the B-band luminosity
function of quasars at the same redshift can both be matched with
the mass function of dark matter halos.

Suggestive similarities have been found between the history of the
blue luminosity density produced by stars and quasars (Cavaliere
\& Vittorini 1998). The similarity persists and even strengthens
when the history of star formation is compared with that of the
global volume emissivity of Active Galactic Nuclei (Franceschini
et al. 1999). The evolution of the luminosity density of
extragalactic radio sources also appears to remarkably parallel
that of the star formation rate (Dunlop 1998).

If quasars were in place roughly at the same time as galaxies, it
is natural to expect that, given their high bolometric
luminosities, they play an important role in the early evolution
of galaxies (Silk \& Rees 1998; Blandford 1999; Monaco et al.
2000): they may provide crucial feedback for limiting star
formation, for arresting its growth, for inducing or inhibiting
collapse of further density perturbations in the neighborhood and
for ionizing the inter-galactic medium (Blandford 1999).

In turn, the fuelling of quasars is likely to be tightly related
to galaxy evolution. Quasars are widely believed to be powered by
accretion into a massive BHs. As discussed by Cavaliere \&
Vittorini (1998, 2000) the data on quasar evolution can be
interpreted in terms of fuelling of BHs during the formation of
the host galaxy while, at later times, accretion is rekindled by
interactions of the host galaxy with companions.

It follows that a consistent evolutionary scenario must deal with
the combined evolution of BHs and of their host galaxies. The link
between the two classes of objects is given by the observed
relationship between the BH mass and the mass (or, better, the
velocity dispersion) of the host galaxy (Magorrian et al. 1998;
Ferrarese \& Merritt 2000; Gebhardt et al. 2000). If this
relationship was imprinted during the early phases of the
evolution (Silk \& Rees 1998; Fabian 1999), the formation rate of
spheroids can be derived directly from the formation rate of
quasars, which, in turn, can be inferred from the epoch-dependent
luminosity function, now observationally determined (or at least
estimated) up to $z\simeq 4.5$ (Boyle et al. 2000; Fan et al.
2001), given the quasar lifetime, $\Delta t_Q$. The latter
quantity can be determined from the condition of consistency with
the present day mass function of spheroids (Salucci et al. 1999),
giving $\Delta t_Q \simeq 4\times 10^7\,$yr, well within the range
of current estimates yielding $\Delta t_Q \simeq 8\times
10^6$--$10^8\,$yr (Salucci et al. 1999; Martini \& Weinberg 2001;
Monaco et al. 2000).

In this framework, Granato et al. (2001) propose the following
scenario:

\begin{itemize}

\item Once the effects of cooling and heating processes (the
latter being mostly due to stellar feedback) are properly taken
into account, the timescale for star formation within virialized
dark matter (DM) halos turns out to be relatively short for
massive spheroidal galaxies ($T_{\rm SF}\sim 0.6\,$Gyr for $M_{\rm
DM}\sim 5\times 10^{12} \ M_{\odot}$), while in the case of less
massive halos the feedback from supernovae (and/or from the active
nucleus) slows down the star formation and can also expel a
significant fraction of the gas. {\it The canonical scheme implied
by the hierarchical CDM scenario -- small clumps collapse first --
is therefore reversed when we consider baryons.} During this
phase, large spheroidal galaxies show up as luminous sub-mm
sources, accounting for the $850\,\mu$m counts.

\item When the quasar luminosity reaches a high enough value, its action
(ionization and heating of the gas), together with that of SN
explosions, stops the star formation and eventually expels the
residual gas. This naturally explains the observed correlation
between BH and host spheroidal masses (Silk \& Rees 1998; Fabian
1999). The same mechanism distributes in the IGM a substantial
fraction of metals.

\item A ``quasar phase'' follows, lasting $10^7$--$10^8\,$yr.

\item Intermediate- and low-mass spheroids have lower SFR's and less
extreme optical depths. They show up as LBGs.

\item A long phase of passive evolution follows, galaxy spectra becoming
rapidly red.

\end{itemize}

\noindent Determining the redshift distribution of SCUBA source is
obviously of utmost importance, since it allows us to obtain a
unbiased view of the star-formation history of the universe and
potentially holds the key to understanding the formation and
evolution of the most massive ellipticals. Unfortunately this is a
very difficult job, due to the rather poor angular resolution of
SCUBA and to the faintness of the optical counterparts, making the
identification ambiguous in most cases. Of greatest importance are
deep VLA maps of SCUBA fields which allow us to identify radio
counterparts or place stringent limits on the radio flux of the
sub-mm galaxies. The recent work on sub-mm to radio spectral
indices (Carilli \& Yun 1999, 2000) has provided us with a useful
tool for estimating, or constraining, the redshifts. The redshift
information derived in this manner can be compared with the
spectroscopic redshifts for individual candidate optical
counterparts to determine the reliability of proposed
identifications.

In spite of the fact that only a handful of redshifts of sources
detected in blanck-field SCUBA surveys have been reliably
measured, the results summarized in Dunlop (2001) indicate that
most SCUBA sources are probably at $z>2$, consistent with the
scenario in which the peak AGN emission corresponds to, or even
causes the termination of, major star-formation activity in giant
spheroidal galaxies (Granato et al. 2001). Given that the optical
emission of powerful quasars peaks at $z\simeq 2.5$, dust emission
is expected to peak $\simeq 0.5\,$Gyr earlier, i.e. at $z\simeq 3$
(Dunlop 2001).

Other important implications of this scenario concern {\it
clustering} (Magliocchetti et al. 2001) and {\it lensing}
properties of dusty spheroids (Perrotta et al. 2001b).



At early times, collapsed objects are likely to be in the highest
peaks of the density field, since one needs a dense enough clump
of baryons in order to start forming stars.  Such high-$\sigma$
peaks are highly biased tracers of the underlying density field
(Kaiser 1984; Bardeen et al. 1986; Mo \& White 1996). The bias
factor is strongly dependent on the mass of dark halos. Therefore,
clustering measurements allow estimates of involved masses.
Although the samples of sub-mm galaxies are still too small to
allow determinations of their clustering properties, large
cell-to-cell fluctuations are seen, suggestive of strong
clustering.

Gravitational lensing of extragalactic light by line of sight mass
concentrations can strongly amplify fluxes of distant sources
(Peacock 1982). Although the probability of strong lensing is very
small, its distribution has a power-law tail ($p(A) \propto
A^{-3}$) extending up to very large values ($A_{\rm max} \simeq
10$--30 for extended sources; Perrotta et al. 2001a). If counts
decrease with increasing flux fast enough, the fraction of lensed
sources at bright fluxes may be large. The special properties of
the sub-mm counts make this spectral region ideally suited for
detecting lensed sources. In fact, as noted above, this spectral
region greatly emphasizes high-redshift sources (which have the
highest probability of being gravitationally lensed) and yields
very steep counts, thus maximizing the amplification bias (Peacock
1982; Turner et al. 1984). Thus, we expect that the fraction of
gravitationally lensed galaxies is boosted in large area,
relatively shallow surveys at sub-mm wavelengths (Blain 1996,
1998).

In the scenario by Granato et al. (2001) the bulk of
star-formation in massive early-type galaxies occurred at $z \ge
2$ and was essentially completed at $z \simeq 1$. This translates
into an essentially exponential turnover of the counts, reflecting
the high-mass turnover of the mass (and of the associated
luminosity) function of galactic halos in place at the  relevant
$z$. While this extremely fast decline of the counts at bright
fluxes may imply that unlensed forming spheroidal galaxies will
not be detectable by large area, relatively shallow surveys such
as those to be carried out by ESA's {\sc Planck} satellite, on the
other hand, the amplification bias due to lensing is maximized. As
a result, the counts of this population may be actually dominated,
at the brightest flux levels, by highly magnified sources. This
characteristic feature of the model by Granato et al. (2001) may
help to discriminate between different scenarios, such as the one
proposed by Rowan-Robinson (2001), predicting less steep counts,
implying a much lower fraction of highly magnified, bright sources
(Perrotta et al. 2001b).

\end{document}